\documentclass{article}
\usepackage{spconf,amsmath,graphicx,url,psfrag}

\title{Interference Alignment (IA) and Coordinated Multi-Point (CoMP) with IEEE802.11ac feedback compression: testbed results}

\name{Per Zetterberg
\thanks{This work was done within the framework of the 
Swedish SSF sponsored RAMCOORAN project and the EU
project RAMCOORAN. 
The HiATUS project acknowledges the financial support of the
Future and Emerging Technologies program within FP7 for Research
of the European Commission (FET Open grant number 265578).}}

\address{Access Linnaeus Center\\
KTH Royal Institute of Technology, Osquldas v{\"{a}}g 10,\\
SE-100 44 Stockholm, Sweden, \\
Email: perz@ee.kth.se}

\begin{document}
\ninept
\maketitle
\begin{abstract}
We have implemented interference alignment (IA) and joint transmission coordinated multipoint (CoMP) on a wireless testbed using the feedback compression scheme of the new 802.11ac standard. The performance as a function of the frequency domain granularity
is assessed.
Realistic throughput gains are obtained by probing each 
spatial modulation stream with ten different coding and modulation schemes.
The gain of IA and CoMP over TDMA MIMO is  found to be 26\% and 71\%, 
respectively under stationary conditions.
In our dense indoor office deployment, the frequency domain granularity of the feedback
can be reduced down to every 8th subcarrier (2.5MHz), without sacrificing performance.
\end{abstract}
\section{Introduction}
\label{sec:intro}

Interference alignment (IA) is a theoretically promising scheme
for dramatically increasing the spectrum efficiency of
wireless systems, \cite{CAD:08}.
Prior work on experimentation with IA includes
 \cite{AYA:10,GAR:11,GAR:11b,BRA:12,ZET:12a,MIL:12} and \cite{MAY:13}.
The papers \cite{AYA:10,GAR:11,GAR:11b,ZET:12a} 
consider a scenario with three simultaneous
MIMO $2 \times 2$ inks, while \cite{MAY:13} consider $4 \times 4$
(\cite{MIL:12} considers a $1 \times 1$ case utilizing a techique
called blind IA which is not be considered in this paper).
In the $2 \times 2$ case, IA should ideally provide three spatial
streams while TDMA MIMO can only deliver two, which in principle should give
a throughput gain of around 50\% at high SNR.
The paper \cite{AYA:10} presents results from simulation
on normalized versions of measured channels, and finds the
degrees of freedom of IA and TDMA MIMO to agree
with the theoretical predictions.
The paper \cite{GAR:11} presented the first over-the-air
evaluation of IA with actual data transmission.
The paper presents EVM measurements but bases channel capacity
results on mutual information calculated from channel measurements.
 The paper finds that IA outperforms
TDMA (17bits/s/Hz versus 11bits/s/Hz at the 40\% level of the sum-rate CDF, 
see Figure 11 of \cite{GAR:11}),
but then no MIMO is applied in the TDMA case.
The measurements were performed in a lecture room.
The paper \cite{MAY:13} investigates the degradation of IA due
to channel estimation errors in the same testbed as \cite{GAR:11},
and finds them to be substantial.
In the paper \cite{ZET:12a} we used an OFDM modulation which is
close to the 802.11a/n/ac standards, and presented results from an 
indoor environment in both in LoS and NLoS locations (between rooms). 
The measurement scenario is clearly interference limited and shows fair
agreement between theory and measurements, if dirty RF effects
are catered for.
The paper \cite{MAY:13} considers an urban mixed outdoor-to-indoor 
and indoor-to-indoor scenario. The mobile-station is located indoors on the fifth floor
of an urban building, two of the base-stations on roof-tops
some 150 meters from the mobile, and the third base-station
in the room adjacent to the mobile (there is only a single
mobile-station, the other two are simulated).
In the paper the performance of IA is analyzed by measuring
the singular values
of the interference at the receiver.
Impressive 39dB interference suppression is achieved. However,
only a single subcarrier is used - which should eliminate most nonlinear
effects. 
 
Prior CoMP experiments are found in \cite{JUN:10,DON:11,HOL:11} and \cite{ZET:12a}. 
The paper \cite{JUN:10} presents experimental results from a real-time LTE 
implementation 
with two base-stations (BSs) and two mobile-stations (MSs), all using two antennas.
The paper reports 
4 to 22 times improvement compared to uncoordinated transmissions
in the two cells. 
The paper \cite{DON:11} presents similar real-time implementations
and reports 2.6 times gain over uncoordinated cells 
(see Table II of \cite{DON:11}). 
The paper \cite{HOL:11},
reports 31\% and 55.3\% gains over uncoordinated cells.
In the paper \cite{ZET:12a} we argued that by modeling the error
vector magnitudes (EVM) of the transmitter radios, fair agreement
can be obtained between measured and real-world SINR ratios for CoMP.

Feedback compression in MIMO interference alignment
has been studied in \cite{Krishnamachari2010,Kim2012,Farhadi2011,REZ:12}
and many other papers. 

In this paper, we extend the implementation of \cite{ZET:12a},
with the use of the practical feedback compression
defined in the IEEE802.11ac standard for single BS multi-user
MIMO (MU MIMO), \cite{POR:13}, and use it in the context of both IA and CoMP. The performance is assessed as a function of the frequency
domain granularity of the feedback. 
We also introduce adaptive modulation and coding and thereby
produce highly realistic performance results.

The paper is organized as follows. In Section \ref{overview} 
we give an overview
of the implementation. The IEEE802.11ac single-cell MU-MIMO feedback scheme 
is reviewed
in Section \ref{feedback} where we also show how to apply it for IA and CoMP.
In the section following, the air interface
and the signal processing functions are described in more detail.
The results are presented in Section \ref{res}. The paper
is concluded in Section \ref{conc}.

\section{Implementation}
\label{overview}


In the considered scenario, there are three base- and
mobile-stations each having two vertically polarized antennas. Under IA
all three BSs simultaneously transmits a modulated
stream to its associated MS by means of linear
precoding at the transmitter, and MMSE combining at the receiver.

As a baseline we use TDMA MIMO where each BS
transmits two streams, but only one BS is active at a time.
We also include two additional baselines. We call them full-reuse MIMO 
and full re-use SIMO. Full-reuse MIMO is identical to TDMA MIMO
except all six streams are active all the time. If the interference
between the BSs is small, this technique will be favorable.
In full-reuse SIMO only one stream is transmitted from each 
BS, just as in IA. When only one of the two 
of interfering BSs is interfering (for instance the other 
one is blocked by a brick-wall), this technique will perform similar to IA
since the receiver can reject the single remaining interferer. 

In addition we include a technique which is superior to 
interference alignment, but more complex, namely joint 
transmission coordinated multipoint 
(CoMP). In this technique all the six transmitter antennas act as one 
BS with six distributed antennas. This technique requires all user 
data to be shared among the three BSs and phase coherence 
among the BS, which is not required in IA.
Both IA and CoMP require precise time alignment 
and coordinated scheduling of transmissions
between the BSs. The precise time alignment enables IA and CoMP to treat 
each subcarrier as an independent narrow-band system

In CoMP we use three streams (one per BS) just as in IA. Adding more streams
could further increase the capacity of CoMP. This is a topic
for future investigation.
%
All the software is available
at \url{sourceforge.net} under a GNU open source software license.
The code can be found under the project names {\tt fourmulti} and 
{\tt iacomp}.
The base-band processing of all BSs are done in one PC and
the corresponding MSs processing in another.
The individual nodes run as separate threads in the PCs.
For a description of the hardware we refer the reader to \cite{ZET:12a}.

The system has been calibrated so that the noise standard deviation 
is roughly the same in all receiver chains, $\sigma^2_{\text{nominal}}$.
The value $\sigma^2_{\text{nominal}}$ is known by all nodes and is used
in the receiver MMSE algorithm, in the feedback quantization, and the 
beamforming max-SINR algorithm.

\begin{figure}
\centerline{
   \includegraphics[width=0.36\textwidth, height=0.225\textheight ]%
     {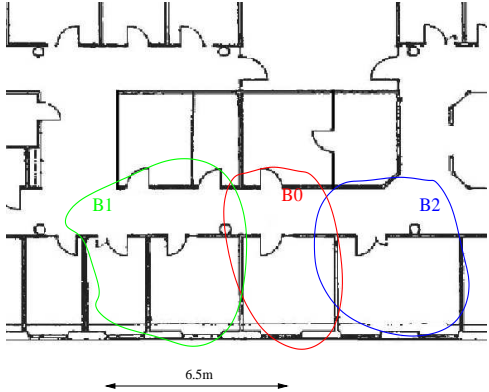}}
\caption{Map over the measurement area}
\label{B0}
\end{figure}

%
The three BSs are placed near the ceiling
in the positions indicated as B0, B1 and B2, in the map
in Figure \ref{B0}. The areas where the corresponding
MS is located during the measurements
are marked with red, green and blue color, respectively.
The LoS measurements were made in the corridor while the 
NLoS measurements were made in the adjacent rooms.

\section{Air Interface and Signal Processing Overview}
\label{air}

The air interface uses OFDM modulation with the same subcarrier
spacing as 802.11a/n/ac namely 312.5kHz. Due to implementation constraints,
we use only 38 subcarriers instead of the (at least) 58 used by 802.11n/ac. 
A cyclic prefix of 0.4$\mu$s is employed, which correspond
to the short cyclic prefix of 802.11n/ac. 

A coding and modulation gearbox with ten combinations of QAM modulations
QPSK, 16QAM, 64QAM and 256QAM and LDPC codes with 1/2, 5/8 and 3/4 rate has
been implemented. The performance of the gearbox is 2-5dB from the 
Shannon limit on AWGN channels in the 2 to 22dB SNR range.

\begin{figure}
\centerline{
   \includegraphics[width=0.45\textwidth, height=0.05\textheight ]%
     {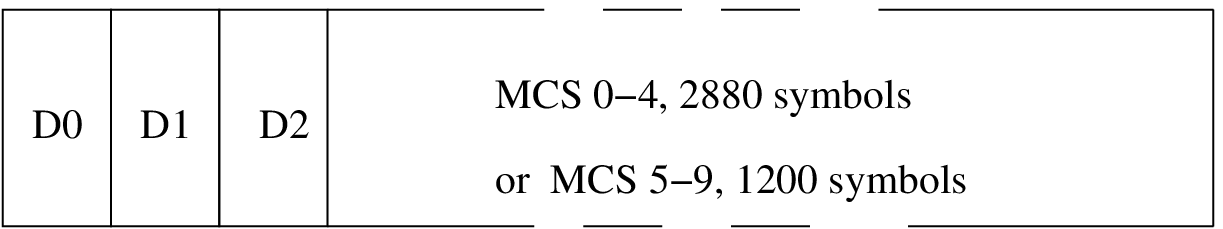}}
\caption{Training block}
\label{frame}
\end{figure}

When running the system, two frames are transmitted. The first frame
contains six OFDM training symbols from the six transmitter antennas
in the system. These pilots model the null data packet (NDP) of IEEE802.11ac,
 see \cite{POR:13}.
The MSs estimates the channels based on these six
OFDM symbols for all six transmitter antennas and all 38 subcarriers.
These estimates are then compressed as described in Section \ref{feedback}.
The result is then sent over wired Ethernet to the transmitter PC. 
In the BS PC,
the channel state information is collected in the
master thread, which calculates the beamforming weights
according the max-SINR algorithm as described in Section \ref{maxSINR}.
A second frame is then formatted and transmitted from all three 
BSs 
simultaneously 40ms after the training frame was transmitted
(the measurements are done under stationary conditions).
This frame is 3.2ms long 
and is divided into six identical subframes 
(for measurement purposes).
Each subframe is divided into two so-called training blocks. A training block
is formatted as indicated in Figure \ref{frame} above.
The three symbols, D0, D1 and D2, are known pilot symbols 
pre-coded with the beamformer
of stream 0,1 and 2, respectively. These symbols correspond to demodulation reference
symbols in LTE nomenclature and correspond to the VHT-LTF field in 802.11ac.
The MMSE receiver uses these three symbols to calculate the combiner weights
i.e. a structured covariance matrix estimate is used. No frequency 
domain averaging of the channel estimates are done.
Following these three training symbols are coding blocks 
which use  MCS0 (MCS stands for coding and modulation scheme)
to MSC4 in the first training block and MCS5 to MCS9 in the second.
The reason for splitting the subframes into two training blocks,
is the channel and sample-clock drifts that need to be compensated by the 
receiver. Tracking loops could maybe eliminate the need for the split.
Before sending the combined samples to the detector, the effective
noise variance is calculated by measuring the distance from the samples
to the nearest constellation points. This noise variance estimate
is used by the log-likelihood ratio (LLR) extractor. This
re-estimation of the noise variance is done in order to account
for the influence of distortions and residual interference.


\section{Feedback according to IEEE802.11ac}
\label{feedback}

The feedback described in standard IEEE802.11ac seems
to originate with \cite{ROH:07} where it is called
``simple feedback method for slowly time-varying channels''.
In this scheme a singular value decomposition
of the MIMO channel (for a certain subcarrier) 
is first performed as ${\bf H} = {\bf U} {\bf S}{\bf V}^H$.
%
In single-user MIMO and single-cell multi-user MIMO, the
${\bf U}$ matrix, is of little use for the transmitter
and therefore only the diagonal matrix ${\bf S}$
and the unitary matrix ${\bf V}$ need to be feed-back 
to the transmitter. When IA and CoMP are considered, several
channel matrices are involved for each user. In this
context the ${\bf U}$ matrix will be different for different links,
and will be important to determine the interference subspaces
at the receivers.
To solve this problem, each MS will instead
base its feed-back on 
the ``big'' ${\bf H}$ matrix, where the submatrices
of the BSs have been concatenated. For instance the
"big" ${\bf H}$ matrix of user $k$ is given by

\begin{equation}
{\bf H}_k = [{\bf H}_{k,1},{\bf H}_{k,2},{\bf H}_{k,3}],
\end{equation}
where ${\bf H}_{k,n}$ is the channel matrix between user $k$ and BS $n$. The MS makes an SVD of this matrix as 

\begin{equation}
{\bf H}_k = {\bf U}_k {\bf S}_k {\bf V}^H_k.
\end{equation}
Since all signals received by the $k$th user will pass through ${\bf H}_k$, it is clear that ${\bf U}_k$ can be neglected. Thus, once the base-stations receive ${\bf U}_k$ and ${\bf S}_k$, they re-create the channel matrix as 

\begin{equation}
\breve{\bf H}_k = {\bf S}_k{\bf V}_k^H,
\label{breveh}
\end{equation}
from which the sub-matrices corresponding to different BSs
can be extracted and applied in the max-SINR algorithm, 
see Section \ref{maxSINR}.

The 802.11ac feedback compression starts 
by phase rotating the columns of ${\bf V}$ so that the last row becomes real 
and positive. These
phases do not need to be sent to the BS 
since demodulation reference signals are used to compensate for
phase rotations of signal streams anyway. In the next step, 
the ${\bf V}$ matrix is multiplied
by a diagonal matrix as

\begin{equation}
\tilde{\bf V} \leftarrow \text{diag} (\exp(j\phi_{1,1}),\ldots,\exp(j\phi_{m-1,1}),1) 
{\bf V}
\end{equation}
where the angles $\phi$ are chosen to remove the phases
of the first column of ${\bf V}$.
These angles $\phi$ are uniformly 
quantized with $b_{\phi}$ bits.
Following this step, is a step where real-valued Givens rotations
are applied to successively zero out the element (2,1) to (m,1)
of $\tilde{\bf V}$ using angles
$\psi_{2,1},\ldots,\psi_{m,1}$. These $\psi$ lie
 between $0$ and $\pi/2$ and are quantized uniformly with 
$b_{\psi}$ bits. The process continues in a similar familiar fashion for the remaining columns of ${\bf V}$.
For the details we refer
the reader to the Matlab/Octave functions available at 
\url{http://people.kth.se/~perz/packV/},
and to \cite{ROH:07} and \cite{POR:13}.
The total number of bits needed to feedback the ${\bf V}$
matrix is given by $((2m-1)n-n^2)(b_{\phi} + b_{\psi})/2$.

In the IA case (with three BSs and MSs) we can actually reduce the number of bits by $2 b_{\phi}$
by reducing the number of $\phi$ angles. This is done by dividing
the ${\bf V}$ matrix in three parts as 

\begin{equation}
{\bf V}^T= [{\bf V}_0^T,{\bf V}_1^T,{\bf V}_2^T].
\end{equation}
Since the signals transmitted from BS0, BS1 and BS2 only propagates through 
${\bf V}_0$, ${\bf V}_1$ and ${\bf V}_2$, respectively,
the beamforming effect is not changed if these blocks are 
rotated by a complex phasor. Thus we modify ${\bf V}_0$ and ${\bf V}_1$ 
so that their upper-left element becomes positive and real-valued.
This will make the corresponding $\phi$ values zero by definition and don't
need to be feed back to the transmitter. In the implementations presented
in the paper, this reduction of the number of feedback bits for IA 
has been implemented.

Since OFDM is used, there
is one ${\bf H}$ matrix per subcarrier. In the 802.11ac standard there is
a parameter $N_g$ which determines the frequency domain granularity of the
feedback. If $N_g=1$ then the feedback of ${\bf V}$ is done on every
subcarrier, if $N_g=2$ the feedback is done on roughly every other subcarrier
and so on. The settings $N_g=1,2,4$ are defined in the standard. We have 
augmented this with $8,16,38$ since this seems to be an effective way
of reducing the number of feedback bits. 
The number of bits for the angles $b_{\phi}$ and $b_{\psi}$ (see above),
can have the values $b_{\psi}=5$ and $b_{\phi}=7$ or 
$b_{\psi}=7$ and $b_{\phi}=9$ according to the standard. Herein,
only the latter values has been used.

The signal to noise ratio (SNR) is reported
with half the granularity of the ${\bf V}$ matrix i.e., 
the SNR is only reported for half
of the subcarriers on which the ${\bf V}$ matrix is reported.
The SNR is reported as corresponding to the diagonal elements of the
matrix ${\bf S}$ (called streams in 802.11ac). In our case we have divided the elements with
noise standard deviation $\sigma_{\text{nominal}}$ mentioned in Section \ref{overview}.
The reporting is done in two steps. First the
SNR averaged (in dB) over the whole band (per singular value)
is reported.
We interpret this as the average over the subcarriers for which the SNR is 
reported.
This average is then uniformly quantized with eight bits in the range
from -10dB to 53.75dB. 
Having obtained the average SNR per stream, the SNR per reported
subcarrier is given as the difference (the delta) between the subcarrier
SNR and its average (in dB). The delta is reported as an integer between
-8dB and +7dB, thus requiring four bits. 

In the BS, we first reconstruct the SNR per reported subcarrier. We then
employ linear interpolation (in dB) to obtain the SNR for all the subcarriers
where ${\bf V}$ is feed back. With these two entities at hand,
the channel matrices are reconstructed as in (\ref{breveh}).
The beamforming weights are calculated as described in Section \ref{maxSINR} based
on these channel matrices. 
On subcarriers where there is no feedback available,
the beamformer of the nearest reported subcarrier is used.
Thus no interpolation is attempted.

\section{Beamformer Implementation}
\label{maxSINR}
%
%
%


We use the max-SINR algorithm of \cite{GOM:08} to iteratively maximize 
the SINR  with respect to the receiver- and transmitter-filters in both the IA and CoMP case.
Compared to the max-SINR implementation in \cite{ZET:12a} two improvements
have been made. First, we initialize the transmitter filters in the closed
form solution in Appendix II of
\cite{CAD:08} for IA. For CoMP we initialize the transmitter filters
in the pseudo-inverse of the eigenbeamformers. Secondly, we have added
a regularization term to the noise as

\begin{equation}
\sigma^2_k = \sigma^2_{\text{nominal}} + \mu \sum_j \| {\bf H}_{k,j} \|^2,
\label{loading}
\end{equation} 

with $\mu=0.001$ to help robustify the system. Equation is based on
the assumption that hardware impairments add a noise proportional 
to the total received power.

\section{Results}
\label{res}

The system was run during night-time with no people moving in the environment. 
In total 22 LoS and 21 NLoS positions were selected for each of the three
MSs. There were several wavelengths between these positions.
For all these positions IA and CoMP and all the reference schemes
were run in a sequence with an interval of 0.34 seconds between the schemes.
For each run we get six throughput values 
by finding the highest MCS with
no bit errors, see Section \ref{air}.
Six seconds later, all the schemes was run again in the same position
with a different value of $N_g$, see Section \ref{feedback}. 
Finally, we also test the performance without quantization
but with frequency domain granularity. 
The reference schemes were repeated for each value of $N_g$,
to ``calibrate'' the throughput gains against the possible influence of 
channel fluctuations.
The average signal to interference ratio (disregarding beamforming)
is 3.2dB. 
The signal to noise ratio (averaged over antennas and subcarriers)
varies between 35 and 60dB in the measurements - thus
the system is truly interference limited.

\begin{figure}
\begin{psfrags}
\psfrag{21}[c][c][0.7]{1}
\psfrag{21.5}{}
\psfrag{22}[c][c][0.7]{2}
\psfrag{22.5}{}
\psfrag{23}[c][c][0.7]{4}
\psfrag{23.5}{}
\psfrag{24}[c][c][0.7]{8}
\psfrag{24.5}{}
\psfrag{25}[c][c][0.7]{16}
\psfrag{25.5}{}
\psfrag{26}[c][c][0.7]{38}
\psfrag{Ng}[c][c][0.7]{$N_g$}
\centerline{
   \includegraphics[width=0.45\textwidth, height=0.18\textheight ]%
     {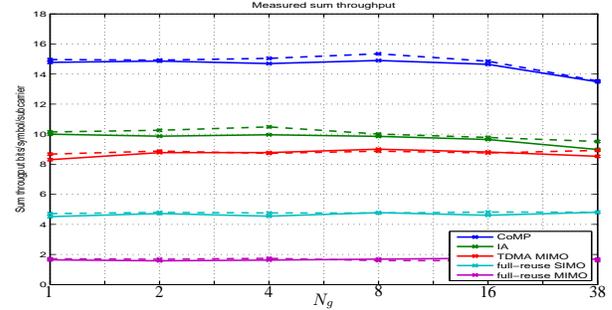}}
\end{psfrags}
\caption{Measured sum throughput in LoS scenarios. The dotted
line are the results without compression}
\label{res1a}
\end{figure}

\begin{figure}
\begin{psfrags}
\psfrag{21}[c][c][0.7]{1}
\psfrag{21.5}{}
\psfrag{22}[c][c][0.7]{2}
\psfrag{22.5}{}
\psfrag{23}[c][c][0.7]{4}
\psfrag{23.5}{}
\psfrag{24}[c][c][0.7]{8}
\psfrag{24.5}{}
\psfrag{25}[c][c][0.7]{16}
\psfrag{25.5}{}
\psfrag{26}[c][c][0.7]{38}
\psfrag{Ng}[c][c][0.7]{$N_g$}
\centerline{
   \includegraphics[width=0.45\textwidth, height=0.18\textheight ]%
     {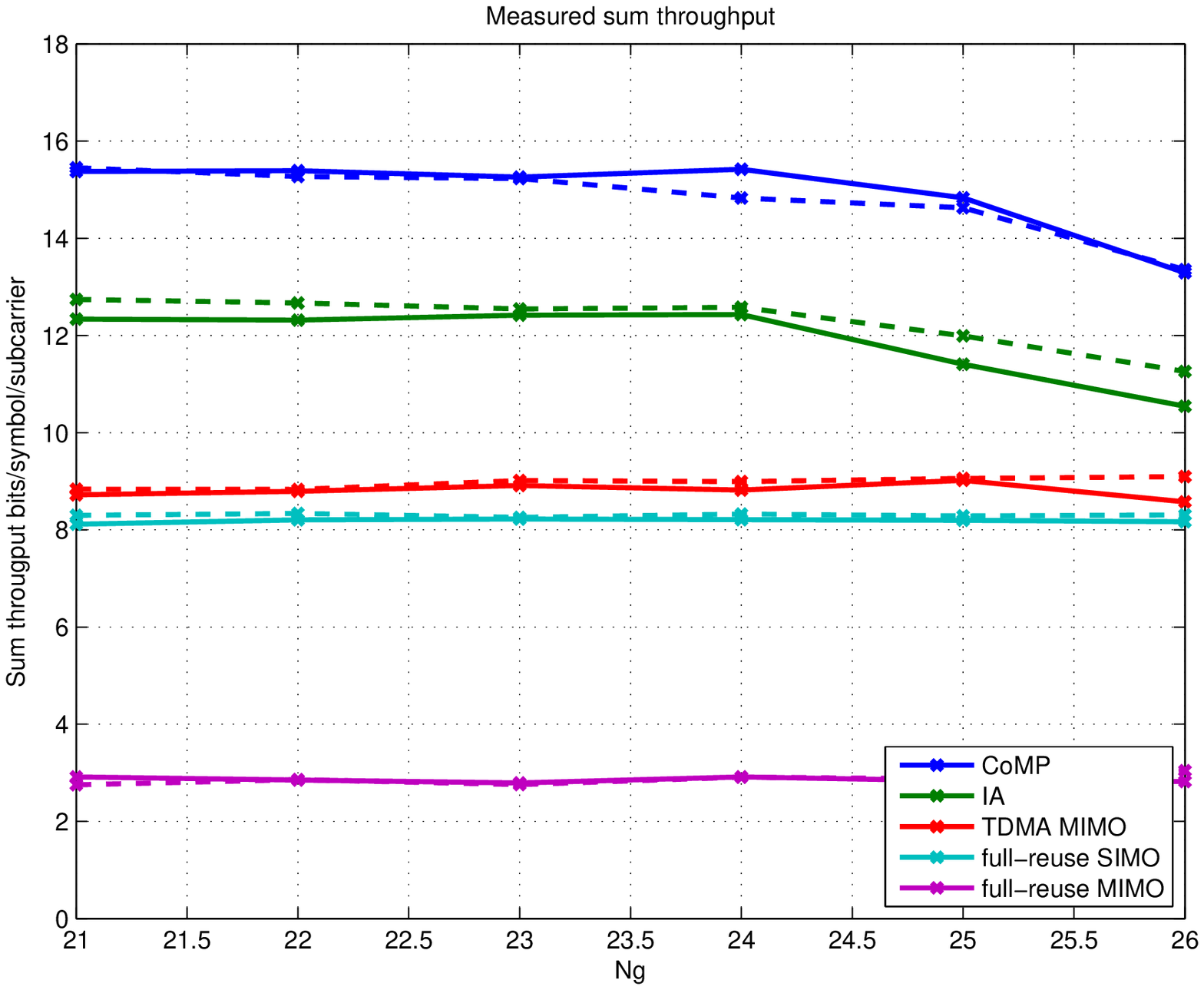}}
\end{psfrags}
\caption{Measured sum throughput in NLoS scenarios. The dotted
line are the results without compression}
\label{res1b}
\end{figure}

\begin{figure}
\begin{psfrags}
\centerline{
   \includegraphics[width=0.45\textwidth, height=0.18\textheight ]%
     {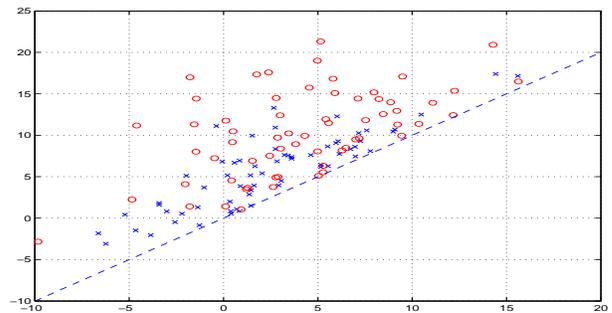}}
\end{psfrags}
\caption{The C/Imin versus C/Imax. 'X': LoS, 'o':NLoS}
\label{res3}
\end{figure}

Figure \ref{res1a} and \ref{res1b} shows the results
from LoS and NLoS, respectively.
We note that the results in LoS and NLoS are similar
for CoMP, and TDMA-MIMO while the results for
IA, full-reuse SIMO and full-reuse MIMO, have improved
by 25\%, 77\% and 76\%, when going from LoS to NLoS respectively.
The reason for this difference can be seen in Figure \ref{res3}.
In Figure \ref{res3} each 'x' represents one mobile LoS position.
Since there are two interfering
BSs, we can define two C/I for each point.
The x-axis of each point represents the lower of the two
C/I and the y-axis the higher. Thus by construction each
'x' is located above the blue dashed $y=x$ line.
The red 'o' points indicates the corresponding statistic
for the NLoS points. From Figure \ref{res3} we see that the C/I
is higher in NLoS than in LoS. Moreover, in LoS it common
that one of the interfering BS is much stronger than the other.
This is probably due to the influence of the walls.
The schemes IA, full-reuse SIMO and full-reuse MIMO, benefit    
from the higher C/I. Full-reuse SIMO further benefit when only
one interfering BS is strong, since it is able to reject
one interferer using its two receive antennas.

From the results we conclude that there is no loss of performance due to the
quantization. 
In LoS  only the lowest frequency domain granularity implemented,
$N_g=38$ results in a loss significant loss while $N_g=16$ incurs
a minor loss. In NLoS a loss of 8\% and 3\%  occurs at $Ng=8$
for IA and CoMP, respectively. For $Ng=16$ the loss increases
to 15\% for both schemes.
. By averaging all measurements where $N_g<16$,
we obtain the following throughputs averaged over
LoS and NLoS
(in sum throughput bits/symbol/subcarrier) IA:11.1 , CoMP:15.1, 
TDMA-MIMO:8.8, full-reuse SIMO:6.4 , full-reuse MIMO:2.2.

\begin{figure}
\begin{psfrags}
\psfrag{21}[c][c][0.5]{1}
\psfrag{21.5}{}
\psfrag{22}[c][c][0.5]{2}
\psfrag{22.5}{}
\psfrag{23}[c][c][0.5]{4}
\psfrag{23.5}{}
\psfrag{24}[c][c][0.5]{8}
\psfrag{24.5}{}
\psfrag{25}[c][c][0.5]{16}
\psfrag{25.5}{}
\psfrag{26}[c][c][0.5]{38}
\psfrag{Ng}[c][c][0.7]{$N_g$}
\centerline{
   \includegraphics[width=0.45\textwidth, height=0.18\textheight ]%
     {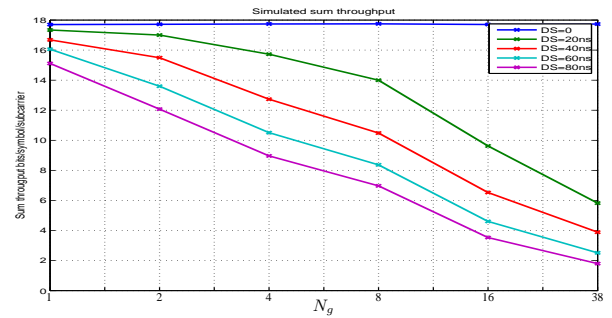}}
\end{psfrags}
\caption{Simulated sum throughput for IA as a function of $N_g$}
\label{res2}
\end{figure}

In Figure \ref{res2} we present simulation results obtained for IA using 
the exact same
C++ code as was running in our real system, but using propagation models
with various levels of RMS delay-spread, representative
for indoor WLAN, \cite{VER:04}.
We first notice that the 
maximum throughputs are higher than in the real system. This is because
the real system is suffering from distortions, see 
\cite{ZET:12a}. The simulated performance degrades quickly when $N_g$ is 
increased. It is possible that some of this effect is masked by
the distortions in the real system. However, the simulation results show
that $N_g<=4$ is necessary for typical conventional deployments.
However, our measurements shows that $N_g$ values greater than four 
would be useful
for very dense deployments. Identical simulations for CoMP
showed less sensitivity to $N_g$.

\section{Conclusion}
\label{conc}

We have implemented 802.11ac based feedback compression in a real system
employing three BS and three MS.
The throughput has been assessed by probing the spatial
channels with a sequence of coding and modulation schemes.
The gain of IA and CoMP over TDMA MIMO is 30\% and 70\% , respectively
under stationary conditions. 
In our dense deployment, the frequency domain granularity of
the feedback can be reduced 
down to about every 8th subcarrier (5MHz), without sacrificing performance.
The feedback compression of 802.11ac does not incur any performance loss.
 
In order to estimate the overhead loss of IA and CoMP under non-stationary 
conditions, the required update
rate need to be known. This will be a topic for future work.

\newpage
\bibliographystyle{IEEEbib}
\bibliography{/home/perz/iwssip2012/ref.bib}

\end{document}